\documentclass[preprint, double-spaced]{revtex4}
\usepackage{graphicx}
\usepackage{amsmath,bbm}
\usepackage{amssymb,bm}

\begin{document}

\title{QCD Critical Point in a Quasiparticle Model}
\author{P.~K.~Srivastava}
\author{S.~K.~Tiwari}
\author{C.~P.~Singh\footnote{corresponding author: $cpsingh_{-}bhu@yahoo.co.in$}}

\affiliation{Department of Physics, Banaras Hindu University, 
Varanasi 221005, INDIA}

\begin{abstract}
\noindent
Recent theoretical investigations have unveiled a rich structure in the quantum chromodynamics (QCD) phase diagram which consists of quark gluon plasma (QGP) and the hadronic phases but also supports the existence of a cross-over transition ending at a critical end point (CEP). We find a too large variation in determination of the coordinates of the CEP in the temperature (T), baryon chemical potential ($\mu_{B}$) plane and, therefore, its identification in the current heavy-ion experiments becomes debatable. Here we use an equation of state (EOS) for a deconfined QGP using a thermodynamically consistent quasiparticle model involving quarks and gluons having thermal masses. We further use a thermodynamically consistent excluded volume model for the hadron gas (HG) which was recently proposed by us. Using these equations of state, a first order deconfining phase transition is constructed using Gibbs' criteria. This leads to an interesting finding that the phase transition line ends at a critical point (CEP) beyond which a cross-over region exists. Using our thermal HG model, we obtain a chemical freeze out curve and we find that the CEP lies in close proximity to this curve as proposed by some authors. The coordinates of CEP are found to lie within the reach of RHIC experiment. 
\\

 PACS numbers: 12.38.Mh, 12.38.Gc, 25.75.Nq, 24.10.Pa

\end{abstract}
\maketitle 
\section{ Introduction}

\noindent
QCD, the non-abelian gauge theory of strong interaction, predicts a phase transition between a hot and dense hadron gas (HG) and the quark-gluon dominated phase which is called as quark gluon plasma (QGP) [1,2]. However, even after an intensive experimental as well as theoretical research spreading over the last two decades, our knowledge regarding the properties and signals of QGP is still very limited [3,4]. Even the phase boundary between the two phases remains in the literature as a conjectured one because the non-perturbative aspects of QCD are still dominant near the region close to the phase transition. In order to test the appearance of a deconfined QGP which exists in a transient phase, we need a proper understanding of its subsequent hadronization process about which our knowledge is really very poor. The numerical method of lattice QCD can properly describe both the phases i.e., QGP and HG. However, lattice QCD studies have yielded results for finite and large values of temperature T and $\mu_{B}$=0 and now we have surmounted difficulties in getting results for small, non vanishing values of $\mu_{B}$ [5,6]. Thus we feel an urgent need to formulate a phenomenological model which can successfully reproduce the lattice QCD data and hence we can further use it to obtain the properties of QGP and to determine the QCD phase diagram in the entire (T, $\mu_{B}$) plane.\\
In this paper, we present a thermodynamically self-consistent quasiparticle model of QCD which describes a gas of quasiparticles with effective masses generated through the interactions among its basic constituents. This model has been found to work well above and around critical temperature $T_{c}$. In order to describe the low energy HG phase, we work with our own model which has been found to give a proper description of the hot and dense HG [7]. The investigation of the structure of the QCD phase diagram is one of the most important and challenging topics in the nuclear and particle physics today. The precise determination of the phase boundary between QGP and HG at high temperature T and small $\mu_{B}$ has been a subject of intense research in recent years from experimental as well as theoretical points of view [8-9]. Lattice simulations first revealed that the transition between HG and QGP phase at $\mu_{B}$=0 and large T is a crossover transition and there were further indications that the crossover transition turned into a first order chiral phase transition for nonvanishing and finite values of $\mu_{B}$ [10]. Several attempts have since been made to locate precisely the critical end point (CEP) i.e., an ending point of the first order chiral transition as $\mu_{B}$ decreases [11]. Although the existence of CEP was predicted a long time ago  by a few lattice calculations, the absence of the CEP in the phase diagram was also noticed in some recent lattice calculations [12,13]. Thus the location and the existence of the CEP in the phase diagram is still a matter of debate. Therefore, it is worthwhile to investigate the precise location of the CEP and to determine its properties in detail with the help of various phenomenological models [9]. However, confusion prevails since the coordinates of the CEP in the (T, $\mu_{B}$) plane vary wildly in various models. Moreover, we are still not certain whether the conjectured phase boundary represents the chiral and/or deconfining phase transition line. Various calculations based on lattice QCD and/or effective models work with the basic assumption that the finite $\mu_{B}$ chiral phase transition is first order and hence the ending point of the line should give the CEP which is a second-order phase transition point. However, we know that the chiral symmetry is broken in the colour flavour locked (CFL) region which has extremely large values of $\mu_{B}$. Theoretically, this topic has largely been investigated using several phenomenological models which can give results in the entire (T, $\mu_{B}$) plane, whereas ab initio calculations are still limited to very small values in $\mu_{B}$. However, all these efforts result in a very wide variation in the coordinates of the CEP [11]. In a recent paper, we proposed a new EOS for HG fireball where the geometrical size of the baryons in HG is explicitly incorporated as the excluded volume correction in a thermodynamically consistent manner [7]. Furthermore, we used a bag model EOS for the QGP phase and a first order phase transition is constructed by equating pressures of both the phases using Gibbs' criteria. We thus obtain an interesting result that such a simple picture not only reproduces the entire conjectured phase boundary but the first order phase transition line ends at a CEP beyond which a crossover region persists [7]. The cordinates of CEP as obtained in our calculation are found to be compatible with the prediction of a recent lattice gauge calculation of Gavai and Gupta [14]. Most importantly, we find here a deconfining phase transition in contrast to other calculations where the phase boundary depicts a chiral symmetry restoring phase transition. However, bag model is often a crude description to a gas of weakly interacting QGP and the nonperturbative effect in this model arises from the pressure of the vaccum through the use of a phenomenological bag constant. In the present work, we consider QGP as a system of quasiparticles which are quarks and gluons possessing temperature-dependent masses arising due to vacuum interactions [15-18]. Recently these models are made thermodynamically self consistent by incorporating suitable corrections in two different ways and hence are referred as QPM I and QPM II in the following. Moreover, these descriptions were initially given for the gluon plasma only. We extend both the models for a QGP with finite baryon chemical potential $\mu_{B}$ and we further assume their validity starting from a threshold temperature $T_{0}=100 MeV$ below the critical temperature $T_{c}$ and we adjust the parameters of the models accordingly. We further compare their predictions for the energy density $\epsilon$ and the pressure $\it{p}$ of the QGP with those obtained previously from the lattice simulation approach. Furthermore, we obtain a new EOS for the hot and dense HG as formulated in the previous paper and construct the phase boundary by equating the QGP pressure with that of HG pressure.We thus determine the critical parameters in the Gibbs' construction of a first order deconfining phase transition. We draw the  phase boundary line and the end of the line determines the coordinates of the CEP. We also find the existence of a crossover transition lying beyond CEP. Finally we compare our findings with those obtained in various other models.\\
The plan of the paper runs as follows. There are two types of quasiparticle models which are thermodynamically self consistent. In section II, we describe quasiparticles (QP), their corresponding equation of state and discuss the criterion of thermodynamical consistency. In section III, we describe the first thermodynamically consistent quasiparticle model [15] of Gorenstein and Yang, the dependence of the quark and gluon masses on the temperature and the extension of the model to describe the physics at finite $\mu_{B}$. In section IV, we discuss the second thermodynamically consistent quasiparticle model [16] of Bannur and its extension for the finite $\mu_{B}$. In section V, we give our formulation for an excluded volume model for the hot and dense hadron gas, and we discuss about its thermodynamical consistency. Section VI summarizes our results and discussions.

\section{Quasiparticle Models (QPM)}
\noindent
Quasiparticles are the thermal excitations of the interacting quarks and gluons.The quasiparticle model in QCD is a phenomenological model which is widely used to describe the non-ideal bahaviour of QGP near the phase transition points. It was first proposed by Golviznin and Satz [17] and then by Peshier et. al. [18] to explain the EOS of QGP obtained from lattice gauge simulation of QCD at finite temperature. In quasiparticle models, the system of interacting massless quarks and gluons can be effectively described as an ideal gas of 'massive' noninteracting quasiparticles. The mass of these quasiparticles is temperature-dependent and arises because of the interactions of quarks and gluons with the surrounding matter in the medium. These quasiparticles retain the quantum numbers of the real particles i.e., the quarks and gluons. It was assumed that energy $\omega$ and momentum k of the quasiparticles obey a simple dispersion relation :
\begin{equation}
\omega^2(k,T)=k^2+m^2(T),
\end{equation} 
\noindent
where m(T) is the temperature-dependent mass of the quasiparticle. The pressure and energy density of the ideal gas of quasiparticles are dependent on $\omega$ and $m(T)$ and are given by [15]:
\begin{equation}
 \it {p}_{id}(T,m)=\mp \frac{Td}{2 \pi^2}\int_{0}^{\infty}k^2 dk \ln\left[1\mp exp\left(-\frac{(\omega-\mu)}{T}\right)\right ],
\end{equation}
\begin{equation}
\epsilon_{id}(T,m)=\frac{d}{2 \pi^2}\int_0^\infty k^2 dk \frac{\omega}{\left[exp\left(\frac{\omega-\mu}{T}\right)\mp1\right]} ,
\end{equation}
\noindent
where d represents the degeneracy factor for quarks and/or gluons. However, Gorenstein and Yang pointed out that this model involves a thermodynamical inconsistency [15]. It did not satisfy the fundamental thermodynamic relation: $\epsilon(T)=T\frac{d\it {p}(T)}{dT}-\it {p}(T)$. So they reformulated the statistical mechanics for a system whose constituents follow a medium-dependent dispersion relation and the above inconsistency problem was handled by them by introducing a temperature-dependent vacuum energy term which effectively cancelled the inconsistent term. Alternatively, Bannur also pointed out the reason for the above thermodynamical inconsistency [16]. If the particle mass in the system is not constant and it depends upon the medium, then the relation used between the pressure and grand canonical partition function does not hold good. So one can  start from the definitions of the energy density and the average particle number density in the grand canonical ensemble formalism and in this way a different thermodynamically consistent quasiparticle model for QGP can be obtained.
\section {First Quasiparticle Model (QPM I)}
Gorenstein and Yang initially formulated a thermodynamically-consistent quasiparticle model for a gluon plasma at $\mu_{B}$=0 and later they extended it for the QGP having a finite value of $\mu_{B}$. In this model, the effective mass of the gluon changes with T and $\mu_{B}$ as follows [15]:
\begin{equation}
m_{g}^{2}(T)=\frac{N_c}{6} g^{2}(T) T^{2} \left(1+\frac{N_{f}^{'}}{6}\right),
\end{equation}
\noindent
where $N_c$ represents the number of colours. We have also taken $N_c=3$ in our calculation. And,
\begin{equation}
N_{f}^{'}=N_{f}+\frac{3}{\pi^2}\sum_{f}\frac{\mu_f^2}{T^2}.
\end{equation}
\noindent
Here $N_f$ is the number of flavours of quarks and $\mu_f$ is the quark chemical potential belonging to the flavour f. Similarly the effective mass of a quark of flavour f changes with  T and $\mu_{B}$ as [19]:
\begin{equation}
m_{qf}^2(T)=\frac{g^2(T) T^2}{6}\left(1+\frac{\mu_f^2}{\pi^2 T^2}\right),
\end{equation} 
Here $g^2(T)$ is the QCD running coupling constant. We have taken the following form for $g^2(T)$ [20-21]:
\\
\begin{equation}
\alpha_{S}(T)=\frac{g^{2}(T)}{4 \pi}=\frac{6 \pi}{\left(33-2 N_{f}\right)\ln \left(\frac{T}{\Lambda_{T}}\sqrt{1+a\frac{\mu^2}{T^2}}\right)}
\\
\left(1-\frac{3\left(153-19 N_f \right)}{\left(33-2 N_f\right)^2}\frac{\ln \left(2 \ln \frac{T}{\Lambda_T}\sqrt{1+a\frac{\mu^2}{T^2}} \right)}{\ln \left(\frac{T}{\Lambda_{T}}\sqrt{1+a\frac{\mu^2}{T^2}}\right) }\right)
\end{equation}
\noindent
Where  $\Lambda_{T}$ is the QCD scale-fixing parameter which characterizes the strength of the interaction. We have taken $\Lambda_{T}=115  MeV$ in our calculation. Here parameter a is equal to $\frac{1}{\pi^{2}}$ [22].\\

After reformulating the statistical mechanics and incorporating the additional medium contribution, the pressure {\it p} and energy density $\epsilon$ for a system of quasiparticles can be written in a thermodynamically consistent manner as follows [15,18]:
\begin{equation}
{\it p}(T,m)={\it p}_{id}-B(T,\mu_{B}),
\end{equation}
\begin{equation}
\epsilon(T,m)=\epsilon_{id}+B(T,\mu_{B}).
\end{equation}
\noindent
The first term on the right hand side of both the equations are the standard ideal gas expressions given by Eq (1) and (2), respectively. The second term represents the medium contribution: 
\begin{equation}
B(T,\mu_{B})=\lim_{V\rightarrow\infty}\frac{E_0}{V},
\end{equation}
\noindent
where $E_0$ is the vacuum energy in the absence of quasiparticle excitations or zero point energy. However, this energy term is not a constant but depends upon $\mu_{B}$ and T. The $B(T,\mu_{B})$ can be derived as follows [23]:
\begin{equation}
B(T,\mu_{B})=B_{0}-\frac{d}{4 \pi^2}\int_{T_0}^{T}dT \int_{0}^{\infty}\frac{k^{2}dk}{\omega}\frac{1}{\left[exp\left(\frac{\omega-\mu_{q}}{T}\right)\right]\mp 1},
\end{equation}
\noindent
where $B_0$ is the integration constant i.e., the value of $B(T,\mu_{B})$ at $T=T_0$. The expression for $B(T,\mu_{B})$ in Eq. (11) and the forms of Eq. (8-10) give a compelling evidence that $B(T,\mu_{B})$ may be treated as $T$ and $\mu_{B}$-dependent bag constant term [24]. In our calculation we take $B_{0}^{1/4}=185 MeV$ and $T_{0}=100 MeV$. We have also taken $\mu_{B}=3\mu_{q}$.
\section{Second Quasiparticle model (QPM II)}
The second thermodynamically consistent quasiparticle model is given by Bannur [16]. Bannur first figured out why there exists a thermodynamical incosistency in the quasiparticle description of Peshier et. al. The relation between pressure and grand canonical partition function can not hold good if the particles of the system have a temperature-dependent mass. So he used the definition of average energy and average number of particles and derived all the thermodynamical quantities from them in a consistent manner. In this model, the effective mass of the gluon is the same as given in Eq (4). However, the effective mass of the quarks involves the following relation:
\begin{equation}
m_{q}^{2}=m_{q0}^{2}+\sqrt{2}m_{q0}m_{th}+m_{th}^{2},
\end{equation}
\noindent
were $m_{q0}$ is the rest mass of the quarks. In this calculation, we have used $m_{q0}=8 MeV$ for two light quarks ({\it u,d}), and $m_{q0}=80 MeV$ for strange quark. In the above Eq (12) $m_{th}$ represents the thermal mass of the quarks and it can be written as [23]:
\begin{equation}
m_{th}^{2}(T,\mu)=\frac{N_{c}^{2}-1}{8 N_{c}}\left[T^{2}+\frac{\mu_{q}^{2}}{\pi^2}\right]g^{2}(T),
\end{equation}
\noindent
Taking these values for the effective masses, energy density can be derived from the grand canonical  partition function in a thermodynamically consistent manner and is given as :
\begin{equation}
\epsilon=\frac{T^4}{\pi^2}\sum_{l=1}^{\infty}\frac{1}{l^4}\left[\frac{d_g}{2}\epsilon_{g}(x_{g}l)+(-1)^{l-1}d_{q}cosh(\mu /T)\epsilon(x_{q}l)+(-1)^{l-1}\frac{d_{s}}{2}\epsilon_{s}(x_{s}l)\right],
\end{equation}
\noindent
with $\epsilon_{i}(x_{i}l)=(x_{i}l)^{3}K_{1}(x_{i}l)+3 (x_{i}l)^{2}K_{2}(x_{i}l)$, where $K_1$ and $K_2$ are the modified Bessel functions with $x_{i}=\frac{m_i}{T}$ and index i runs for gluons, up-down quarks q, and strange quark s. Here $d_i$ are the degeneracies associated with the internal degrees of freedom. Now, by using the thermodynamic relation $\epsilon=T\frac{\partial \it {p}}{\partial T}-\it {p}$, pressure of system at $\mu_{q}=0$ can be obtained as:
\begin{equation}
\frac{\it{p}(T,\mu_{q}=0)}{T}=\frac{\it{p}_0}{T_0}+\int_{T_0}^{T}dT \frac{\epsilon(T,\mu_{q}=0)}{T^2},
\end{equation}
\noindent
where $\it{p}_0$ is the pressure at a reference temperature $T_0$. We have used $\it{p}_{0}$=0 at $T_{0}$=100 MeV in our calculation. Using the relation between the number density $n_{q}$ and the grand canonical partition function, we can get the pressure for a system at finite $\mu_{B}$ :
\begin{equation}
\it{p}(T,\mu_{q})=\it{p}(T,0)+\int_{0}^{\mu_{q}}n_{q}d\mu_{q}.
\end{equation}
\noindent
Thus all the thermodynamical quantities can be obtained in a consistent way by using this model.

\section{EOS for a Hadron gas}
There is no deconfinement transition, if the hadron gas consists of point-like particles, and consequently HG pressure is always larger than QGP pressure. Therefore, inclusion of a repulsive interaction between two baryons having a hard-core size reduces the HG pressure and hence it stabilizes the formation of QGP at high baryon densities. Recently we have proposed a thermodynamically consistent excluded volume model for hot and dense hadron gas (HG). In this model, the grand canonical partition function for the HG with full quantum statistics and after incorporating excluded volume correction can be written as [25]:

\begin{equation}
\begin{split}
ln Z_i^{ex} = \frac{g_i}{6 \pi^2 T}\int_{V_i^0}^{V-\sum_{j} N_j V_j^0} dV
\\
\int_0^\infty \frac{k^4 dk}{\sqrt{k^2+m_i^2}} \frac1{[exp\left(\frac{E_i - \mu_i}{T}\right)+1}
\end{split}
\end{equation}
where $g_i$ is the degeneracy factor of ith species of baryons,$E_{i}$ is the energy of the particle ($E_{i}=\sqrt{k^2+m_i^2}$), $V_i^0$ is the eigenvolume of one baryon of ith species and $\sum_{j}N_jV_j^0$ is the total occupied volume and $N_{j}$ represents total number of baryons of jth species.

Now we can write Eq.(17) as:

\begin{equation}
ln Z_i^{ex} = V(1-\sum_jn_j^{ex}V_j^0)I_{i}\lambda_{i},
\end{equation}
where $I_{i}$ represents the integral:
\begin{equation}
I_i=\frac{g_i}{6\pi^2 T}\int_0^\infty \frac{k^4 dk}{\sqrt{k^2+m_i^2}} \frac1{\left[exp(\frac{E_i}{T})+\lambda_i\right]},
\end{equation}
and $\lambda_i = exp(\frac{\mu_i}{T})$ is the fugacity of the particle, $n_j^{ex}$ is the number density of jth type of baryons after excluded volume correction and can be obtained from Eq.(18) as:
\begin{equation}
n_i^{ex} = \frac{\lambda_i}{V}\left(\frac{\partial{ln Z_i^{ex}}}{\partial{\lambda_i}}\right)_{T,V}
\end{equation}
This leads to a transcendental equation as
\begin{equation}
n_i^{ex} = (1-R)I_i\lambda_i-I_i\lambda_i^2\frac{\partial{R}}{\partial{\lambda_i}}+\lambda_i^2(1-R)I_i^{'}
\end{equation}
where $I_{i}^{'}$ is the partial derivative of $I_{i}$ with respect to $\lambda_{i}$and $R=\sum_in_i^{ex}V_i^0$ is the fractional occupied volume. We can write R in an operator equation as follows [7]:
\begin{equation}
R=R_{1}+\hat{\Omega} R
\end{equation}
where $R_{1}=\frac{R^0}{1+R^0}$ with $R^0 = \sum n_i^0V_i^0 + \sum I_i^{'}V_i^0\lambda_i^2$; $n_i^0$ is the density of pointlike baryons of ith species and the operator $\hat{\Omega}$ has the form :
\begin{equation}
\hat{\Omega} = -\frac{1}{1+R^0}\sum_i n_i^0V_i^0\lambda_i\frac{\partial}{\partial{\lambda_i}}
\end{equation}
Using Neumann iteration method and retaining the series upto $\hat{\Omega}^2$ term, we get
\begin{equation}
R=R_{1}+\hat{\Omega}R_{1} +\hat{\Omega}^{2}R_{1}
\end{equation}
\noindent
Eq.(24) can be solved numerically. Finally, we get for the total pressure [25] of the hadron gas:
\begin{equation}
\it{p}_{HG}^{ex} = T(1-R)\sum_iI_i\lambda_i + \sum_i\it{P}_i^{meson}
\end{equation}

In (25), the first term represents the pressure due to all types of baryons where excluded volume correction is incorporated and the second term gives the total pressure due to all mesons in HG having a pointlike size. This makes it clear that we consider the hard-core repulsion arising between two baryons which possess a hard-core size. In this calculation, we have taken an equal volume $V^{0}=\frac{4 \pi r^3}{3}$ for each type of baryon with a hard-core radius $r=0.8 fm$. We have taken all baryons and mesons and their resonances having masses upto $2 GeV/c^{2}$ in our calculation for HG pressure. We have also used the condition of strangeness conservation by putting $\sum_{i}S_{i}(n_{i}^{s}-\bar{n}_{i}^{s})=0$, where $S_{i}$ is the strangeness quantum number of the ith hadron, and $n_{i}^{s}(\bar{n}_{i}^{s})$ is the strange (anti-strange) hadron density, respectively. Using this constraint equation, we get the value of strange chemical potential in terms of $\mu_{B}$. We want to stress here that the form of this model used under Boltzmann approximation has been found to describe [26-27] very well the observed multiplicities and the ratios of the particles in heavy-ion collisions.\section{Results and Discussion}
In order to demonstrate that both types of quasiparticle models reproduce the lattice results with the value of parameters chosen here, we show in Fig. 1, the results of our calculations for the variations of energy density with respect to temperature at different $\mu_{B}$. We find that the predictions from both these models (QPM I and QPM II) compare well with the lattice data [28], although it cannot be regarded as a very good fit to the data. However, the results of QPM may improve if we use thermal gluon mass as $m_{g}^{2}(T)=g^{2}(T)T^{2}/3$ instead of $m_{g}^{2}(T)=g^{2}(T)T^{2}/2$ as pointed out by Bannur [16,21]. 

\begin{figure}
\includegraphics[height=28em]{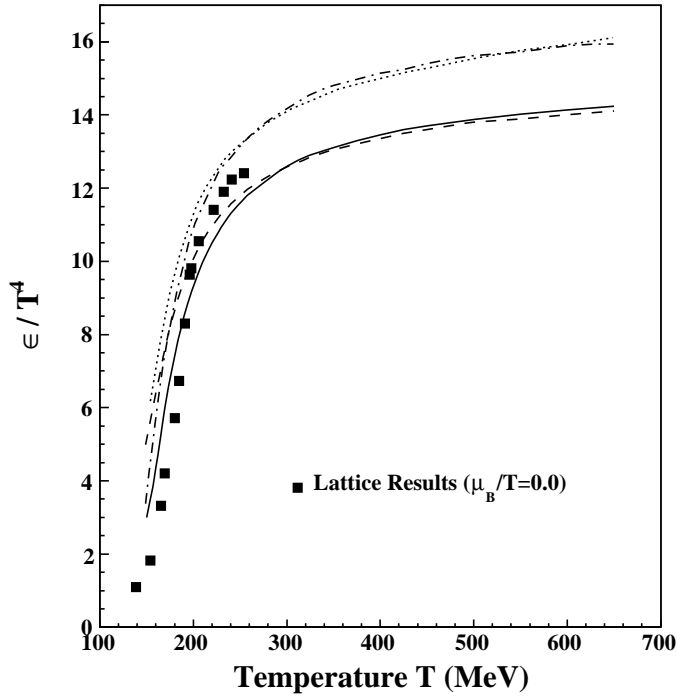}
\caption[]{The calculation of energy density of QGP in QPM I (by a dashed curve at $\mu_{B}$/T=0.0 and by a dotted curve at $\mu_{B}$/T=2.0) and in QPM II (by a solid line at $\mu_{B}$/T=0.0 and by a dashed-dotted curve at $\mu_{B}$/T=2.0). Square points are lattice data at $\mu_{B}$/T=0.0 [28].}
\end{figure}
\begin{figure}
 \includegraphics[height=28em]{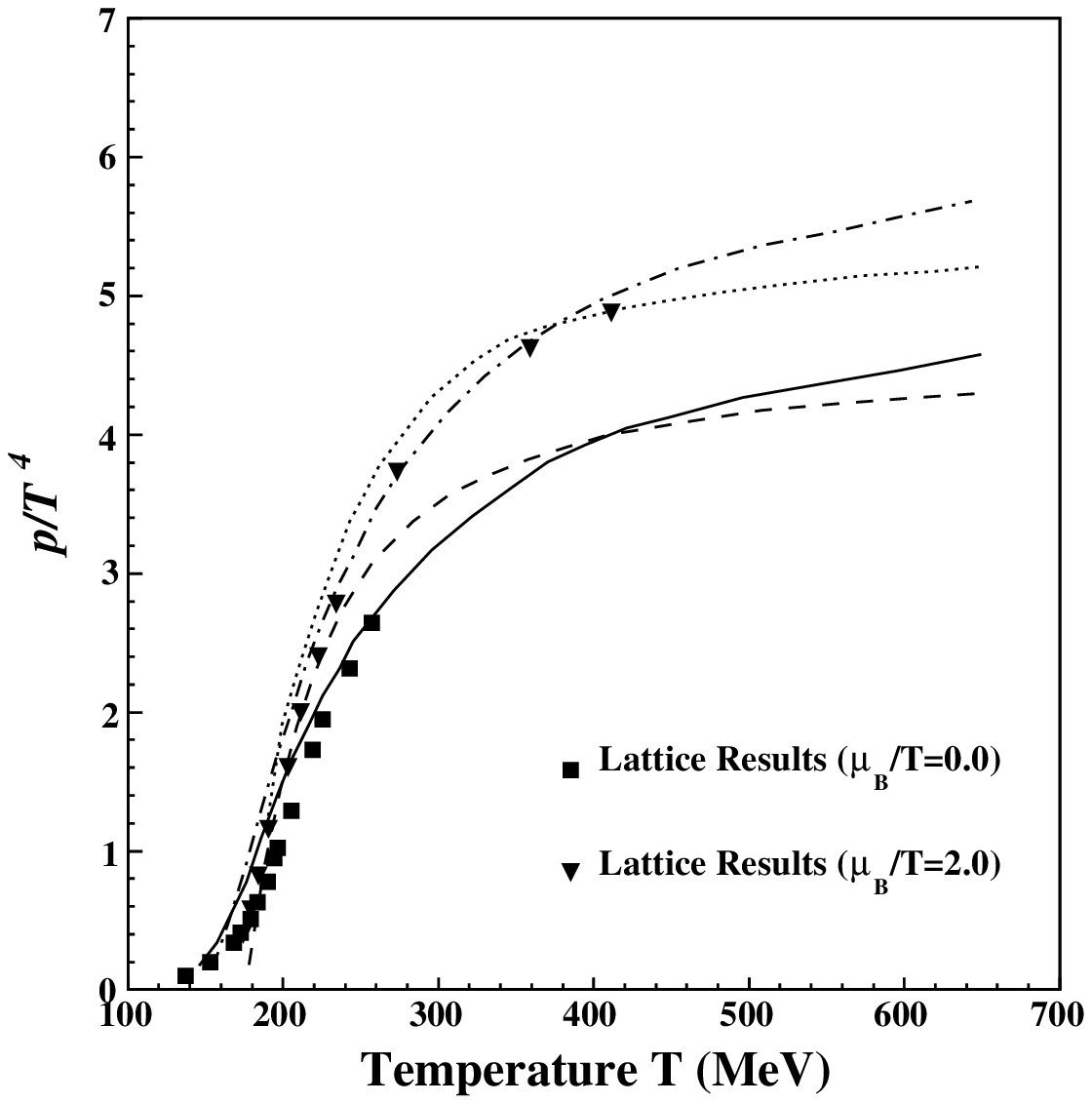}
\caption[]{The variation of QGP pressure with the temperature at different $\mu_{B}$. In QPM I we have shown the results by dashed curve at $\mu_{B}$/T=0.0 and by a dotted curve at $\mu_{B}$/T=2.0. Similarly in QPM II, the solid line represents results at $\mu_{B}$/T=0.0 and the dashed-dotted curve at $\mu_{B}$/T=2.0, respectively. Square and triangular points are lattice data at $\mu_{B}$/T=0.0 and at $\mu_{B}$/T=2.0, respectively [28,29].}
\end{figure}
 
In Fig.2, we have presented the results of our calculations for the QGP pressure $p/T^{4}$ in both the quasiparticle models and shown its temperature variation at different values of $\mu_{B}$. We compare our results with those recently reported in lattice simulations [28,29]. We find that the fits by QPM II look much better than those given by QPM I. These results give us extreme confidence in both types of quasiparticle models being used as phenomenological models. Although phenomenology cannot work as a substitute for a formal theory like QCD. However, since the utility of lattice QCD calculations at very large $\mu_{B}$ is still not possible, we take the help of quasiparticle model which has a few adjustable parameters. It is now widely used to describe the non-ideal behaviour of QGP observed near the critical line. Moreover, we attempt to extend its uses at lower values of temperature, eg., $T_{0}<T_{c}$ (we take $T_{0}=100 MeV$). Usually authors have studied the quasiparticle models above $T_{c}$ only and therefore, the rapid rise of pressure and energy density at or around $T_{c}$ is not properly taken care of in these models. Our method of obtaining critical parameters $(T_{c},\mu_{c})$ involves the use of Gibbs' equilibrium criteria of equating HG and QGP pressures and to determine where these pressure lines intersect each other. Therefore, we want to know precisely the values of QGP pressure at temperatures $T_{0}(<T_{c})$ also. The comparison of our calculations with the lattice results yields the required test about the suitability of quasiparticle EOS for QGP.\\

\begin{figure}
\begin{center}
\noindent
\includegraphics[height=32em]{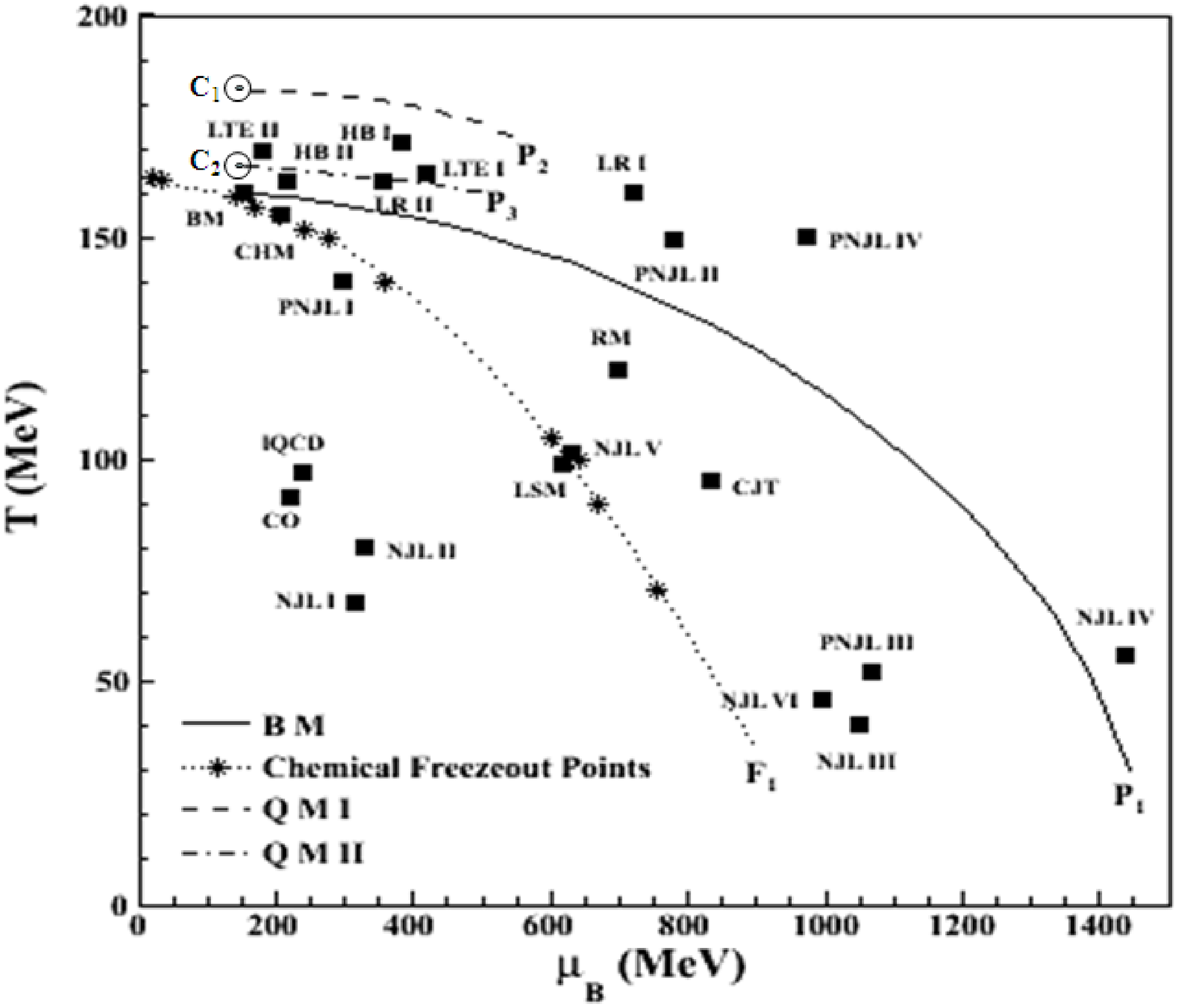}
\caption[]{The location of QCD critical point in QCD phase diagram. $P_{1}$ is the phase boundary in bag model (BM), $P_{2}$ is the phase boundary in QPM I and $P_{3}$ is the phase boundary in QPM II. $F_{1}$ is the chemical freezeout line obtained using our HG model. $C_{1} (T_{c}=183 MeV, \mu_{c}=166 MeV)$ is the CEP on $P_{2}$ obtained in QPM I and $C_{2} (T_{c}=166 MeV, \mu_{c}=155 MeV)$ is the CEP on $P_{3}$ obtained in QPM II. The labels used in the figure are explained in the table I. }
\end{center}
\end{figure}

In Fig.3, we have shown the phase boundary obtained in our model. Surprisingly we again find here that the first-order deconfining phase transition line ends at a critical end point (CEP) and the coordinates of CEP are ($T_{c}=183 MeV, \mu_{c}=166 MeV$) in QPM I and ($T_{c}=166 MeV, \mu_{c}=155 MeV$) in QPM II. It is interesting to find that the critical points obtained by us lie closer to CEP of some lattice calculation [14]. These points also compare well with the coordinates of CEP ($T_{c}=160 MeV, \mu_{c}=156 MeV$) obtained by us in the previous publication [7] using bag model calculation. We also find a crossover region existing beyond the critical point where HG pressure which is solely dominated by mesonic pressure term in Eq.(25), is always less than the QGP pressure. Therefore, no phase transition exists in this region. Since the temperature is much higher, the thermal fluctuations break mesonic constituents of HG into quarks, antiquarks and gluons. We have tabulated in Table I the location of CEP obtained from various calculations for a comparison. For convenience we have shown above the dark solid line, all the values obtained in SU(3) flavour calculations. Below the solid line, the values of SU(2) are also shown in order to make the comparison complete. 
\begin{table}
\includegraphics[height=40em]{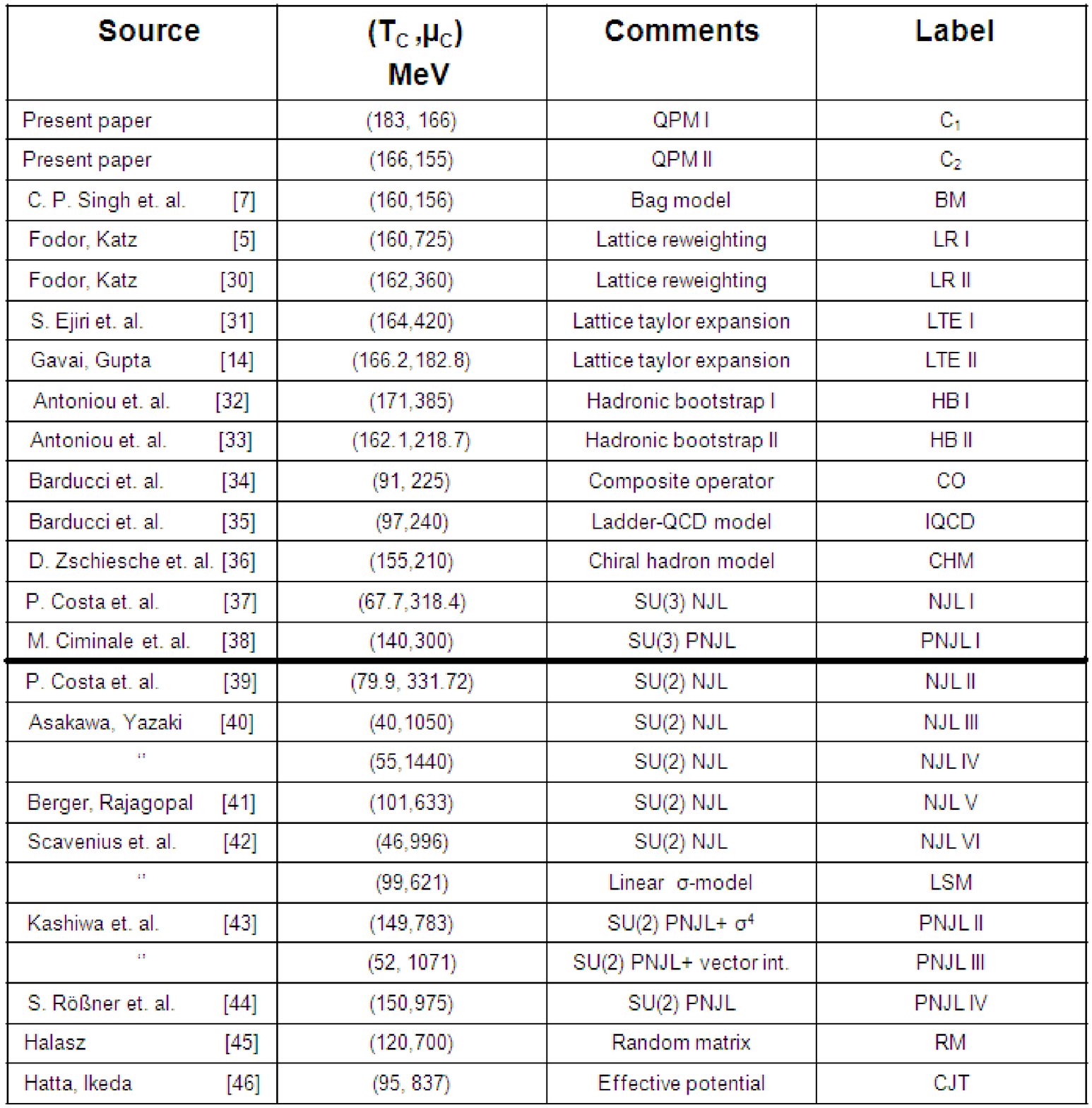}
\caption[]{Coordinates of CEP obtained in different models. The last column gives the corresponding label used in Fig. 3 and Reference [11]}
\end{table}

We find that there exists a very wide variation in the coordinates of CEP obtained in different models. We notice that critical end points obtained in deconfining phase transition are usually located at $\mu_{B}<200 MeV$ whereas chiral CEP have much larger $\mu_{B}$. So there is a good chance for observing CEP at RHIC by using energy scan [47]. We also find that CEP obtained in our models almost overlaps with the points on the freezeout curve. The freezeout point occurs close to RHIC energy and hence the fluctuations in multiplicity etc. can experimentally provide a clear signal for CEP.

Before we conclude we would like to discuss the recent lattice data obtained by de Forcrand and Philipsen [12, 13] who used 2+1 and 3 flavors staggered fermions and a taylor expansion in $\mu_{q}/T$ to study the curvature of the critical surface at very light quark masses close to $\mu_{q}=0$ surface. They noticed that the critical surface bends so that the first order region shrinks at higher quark masses and hence they conclude that there is no critical point at finite chemical potential. However, it is speculated that the critical surface bends back at larger $\mu_{q}$ and the critical point may again reappear. In fact a recent NJL model calculation lends support to this speculation [48].

In summary we have demonstrated the occurrence of CEP in a deconfining first order phase transition constructed by using quasiparticle model for QGP and a new thermodynamically consistent excluded volume model for hot and dense HG. In our model, we assign a hard core volume for baryons and mesons are treated as pointlike particles. So at higher temperatures, mesons can fuse into one another but baryons occupy space. As $\mu_{B}$ increases, we find that the fractional occupied volume R by baryons increases and hence the mobility of baryons decreases fast. The physical mechanism in our model is similar to the percolation model [49] where a first order phase transition results due to $''jamming''$ of baryons in the HG. Thus our finding lends support to the idea of realising a phase transition by modelling the interactions existing in the HG in a suitable way. 

\section {Acknowledgements}
 PKS and SKT are grateful to University Grants Commission (UGC) and Council of Scientific and Industrial Research (CSIR), New Delhi for providing a research fellowship. CPS acknowledges the financial support through a project sanctioned by Department of Science and Technology, Goverment of India, New Delhi.

\pagebreak


\begin{thebibliography}{99}

\bibitem{[1]} C. P. Singh, Phys. Rep. 236, 147 (1993), Int. J. Mod. Phys. A7, 7185 (1992)
\bibitem{[2]} H. Satz, Rep. Prog. Phys. 63, 1511 (2000) 
\bibitem{[3]} E. Shuryak, Prog. Part. Nucl. Phys. 62, 48 (2009)
\bibitem{[4]} H. Satz, Int. J. Mod. Phys A21, 672 (2006)
\bibitem{[5]} Z. Fodor and S. D. Katz, Phys. Lett. B534, 87 (2002); JHEP 0203, 014 (2002)
\bibitem{[6]} C. Schmidt, PoS LAT 2006, 021 (2006)
\bibitem{[7]} C. P. Singh, P. K. Srivastava and S. K. Tiwari, Phys. Rev. D80, 114508 (2009)
\bibitem{[8]} M. A. Stephanov, Phys. Rev. Lett. 102, 032301 (2009)
\bibitem{[9]} M. A. Stephanov, Int. J. Mod. Phys. A20, 4387 (2005)
\bibitem{[10]} M. A. Stephanov, K. Rajagopal and E. V. Shuryak, Phys. Rev. Lett. 81, 4816 (1998)
\bibitem{[11]} M. A. Stephanov, Prog. Theor. Phys. Suppl. 153, 139 (2004)
\bibitem{[12]} P. de Forcrand and O. Philipsen, J. High Energy Phys. 01, 077 (2007)
\bibitem{[13]}P. de Forcrand and O. Philipsen, J. High Energy Phys. 11, 012 (2008)
\bibitem{[14]} R. V. Gavai and S. Gupta, Phys. Rev. D71, 114014 (2005)
\bibitem{[15]} M. I. Gorenstein and S. N. Yang, Phys. Rev. D52, 5206 (1995)
\bibitem{[16]} V. M. Bannur, Phys. Lett. B647, 271 (2007); J. Phys. G: Nucl. Part. Phys. 32, 993 (2006)
\bibitem{[17]} V. Goloviznin and H. Satz, Z. Phys. C57, 671 (1994)
\bibitem{[18]} A. Peshier, B. Kampfer, O. P. Pavlenko and G. Soff, Phys. Rev. D54, 2399 (1996)


\bibitem{[19]} H. Vija and M. H. Thoma, Phys. Lett. B342, 212 (1995)
\bibitem{[20]} E. Braaten and R. D. Pisarski, Phys. Rev. D45, R1827 (1992)
\bibitem{[21]} V. M. Bannur, Eur. Phys. J. C50, 629-634 (2007)
\bibitem{[22]} J. Letessier and J. Rafelski, Phys. Rev. C67, 031902(R) (2003)
\bibitem{[23]} A. Peshier, B. Kampfer and G. Soff, Phys. Rev. C61, 045203 (2000)
\bibitem{[24]} B. K. Patra and C. P. Singh, Phys. Rev. D54, 3551 (1996); N. Prasad and C. P. Singh, Phys. Lett. B501, 92 (2001)
\bibitem{[25]} C. P. Singh, B. K. Patra and K. K. Singh, Phys. Lett. B387, 680 (1996); S.Uddin and   C. P. Singh, Zeit. f. Phys. C63, 147 (1994)
\bibitem{[26]} M. Mishra and C. P. Singh, Phys. Rev. C76, 024908 (2007); Phys. Lett. B651, 119 (2007)
\bibitem{[27]} M. Mishra and C. P. Singh, Phys. Rev. C78, 024910 (2008)
\bibitem{[28]} M. Cheng et. al., Phys. Rev. D81, 054504 (2010); M. Cheng et. al., Phys. Rev. D77, 014511 (2008)
\bibitem{[29]} C. Schmidt, arXiv:0810.4024v1[hep-lat]
\bibitem{[30]} Z. Fodor and S. D. Katz, J. High Energy Phys. 0404, 050 (2004)
\bibitem{[31]} S. Ejiri, C. R. Alton, S. J. Hands, O. Kaczmarek, F. Karsch, E. Laermann and C. Schmidt, Prog. Theor. Phys. Suppl. 153, 118 (2004)
\bibitem{[32]} N. G. Antoniou and A. S. Kapoyannis, Phys. Lett. B563, 165 (2003)
\bibitem{[33]} N. G. Antoniou, F. K. Diakonos and A. S. Kapoyannis, Nucl. Phys. A759, 417 (2005)
\bibitem{[34]} A. Barducci, R. Casalbuoni, G. Pettini and R. Gatto, Phys. Rev. D49, 426 (1994)
\bibitem{[35]} A. Barducci, G. Pettini, L. Ravagli and R. Casalbuoni, Phys. Lett. B564, 217 (2003)
\bibitem{[36]} D. Zschiesche et. al., J. Phys. G: Nucl. Part. Phys. 31, 935 (2005)
\bibitem{[37]} P. Costa, M. C. Ruivo, and C. A. deSousa, Phys. Rev. D77, 096001 (2008)
\bibitem{[38]} M. Ciminale, R. Gatto, N. D. Ippolito, G. Nardulli, and M. Ruggieri, Phys. Rev. D 77, 054023 (2008)
\bibitem{[39]} P. Costa, C. A. deSousa, M. C. Ruivo, and H. Hansen, Europhys. Lett. 86, 31001 (2009)
\bibitem{[40]} M. Asakawa and K. Yazaki, Nucl. Phys. A504, 668 (1989)
\bibitem{[41]} J. Berges and K. Rajagopal, Nucl. Phys. B538, 215 (1999)
\bibitem{[42]} O. Scavenius, A. Mocsy, I. N. Mishustin and D. H. Rischke, Phys. Rev. C64, 045202 (2001)
\bibitem{[43]} K. Kashiwa, H. Kouno, M. Matsuzaki, and M. Yahiro, Phys. Lett. B662, 26 (2008)
\bibitem{[44]} S. Rossner, C. Ratti and W. Weise, Phys. Rev. D75, 034007 (2007)
\bibitem{[45]} M. A. Halasz, A. D. Jackson, R. E. Shrock, M. A. Stephanov, J. J. M. Verbaarschot, Phys. Rev. D58, 096007 (1998)
\bibitem{[46]} Y. Hatta and T. Ikeda, Phys. Rev. D67, 014028 (2003)
\bibitem{[47]} R. A. Lacey et. al., Phys. Rev. Lett 98, 092301 (2007)
\bibitem{[48]} J. W. Chen, K. Fukushima, H. Kohiyama, K. Ohnishi, and U. Raha, Phys. Rev. D80, 054012 (2009)
\bibitem{[49]} P. Castorina, R. Redlich and H. Satz, Eur. Phys. J. C59, 67 (2009)1


\end{thebibliography}
\end{document}